**Sergey Yekimov**

Department of Trade and Finance, Faculty of Economics and Management, Czech University of Life Sciences Prague, Kamycka 129, 16500, Praha - Suchdol, Czech Republic

# USING FERMAT-TORRICELLI POINTS IN ASSESSING INVESTMENT RISKS

## Abstract

The use of Fermat-Torricelli points can be an effective mathematical tool for analyzing numerical series that have a large variance, a pronounced nonlinear trend, or do not have a normal distribution of a random variable.

Linear dependencies are very rare in nature.

Smoothing numerical series by constructing Fermat-Torricelli points reduces the influence of the random component on the final result.

 The presence of a normal distribution of a random variable for numerical series that relate to long time intervals is an exception to the rule rather than an axiom. The external environment (international economic relations, scientific and technological progress, political events) is constantly changing, which in turn, in general, does not give grounds to assert that under these conditions a random variable satisfies the requirements of the Gauss-Markov theorem.

## Introduction

> In every big case, you always have to leave some of it to chance.
> *Napoleon I Bonaparte*

The assessment and analysis of investment projects are based on economic forecasts. The lack and limited access to information, as well as the difficulty of obtaining reliable statistical information, necessitate making investment decisions in conditions of uncertainty and risk. Making an investment decision in conditions of uncertainty can be carried out both on the basis of data on previous activities, and if there was no previous activity. If there was no previous activity, then investment risks can be assessed based on the subjective opinion of experts.

According to [1], different types of risks correlate with each other. When managing investment projects, risk management is required to choose between partially or fully accepting risk, and therefore managing risk, making a decision to avoid or localize risk.

The authors [2] note that investment risks represent a complex of external and internal factors that determine the profitability and effectiveness of the implementation of an investment project. According to [3], the following conditions may accompany an investment decision: the need to choose a solution based on possible alternatives, the need to assess the probability of the origin of certain events, the presence of uncertainty when choosing the right investment decision. According to [4], the following risk groups accompany investment activities: economic, technical, financial, political and monetary. When managing a securities portfolio, investment risks may be associated with an incorrect forecast for prices and sales volumes of securities, an incorrect assessment of investment expectations and the state of the competitive environment.

According to [5], among the circumstances influencing the emergence of investment risks, one can distinguish:

1) Changes in exchange rates and market conditions.
2) The dynamics of social and economic processes.
3) Instability of investment legislation.
4) Lack of free access to information on technical and economic indicators and innovative technologies.
5) The influence of natural and climatic factors.
6) The instability of the political situation.
7) Lack of objective information about the business reputation and financial and economic situation of enterprises involved in investment activities.

The authors [6] functional and operational investment risks. Operational investment risk is associated with the occurrence of failures in information systems, erroneous actions of employees of enterprises, technological errors in the process of carrying out production activities. Functional investment risk is associated with mistakes made in the management and formation of investment portfolios. According to [7], investment risks can be divided into technological (related to production processes), social (related to the influence of the human factor), economic, legal, political and environmental. According to [8], investment risks can be divided into commercial (related to the object of investment) and external risks (related to the influence of external factors that the investor cannot influence). The authors [9] divide investment risks into simple and commercial ones. Business and speculative risks are considered under commercial risks. Under simple risks, the risks associated with the actions of third parties and natural phenomena are considered.

According to [10], investment risks can be divided into static and dynamic. Speculative risks are usually dynamic. Static risks are associated with the occurrence of events, as a result of which further investment activity becomes impractical or impossible. The presence of investment risks necessitates quantitative and qualitative analysis aimed at risk assessment and management. The authors [11] note that a quantitative analysis of the impact of risks on investment activities should involve a numerical measurement of the impact of risk factors on the successful implementation of investment activities.

According to the authors [12], a qualitative assessment of investment risks should be based on the following principles: the independence of the manifestation of investment risk in relation to each specific case, the amount of probable damage from investment activities should not exceed the financial capabilities of investors. According to [13], investment risk is characterized by the presence of:

1. The uncertainty of the occurrence of unforeseen adverse events.
2. The possibility of losses as a result of investment activities.
3. The probabilistic effect of the investment activity.

According to [14], the quantitative assessment of investment risks should include:

1) Studying the impact of an adverse event on the financial results of investment activities.
2) Assessment of the level of risk tolerance relative to the investor's wishes.
3) Analysis of the ratio of the cost of possible losses from investment activities in relation to the costs aimed at reducing possible investment risks.
4) Determination of the probability of occurrence of a risk event for each threat or danger.

According to the authors [15], quantitative assessment of investment risks can be implemented on the basis of the following methods (Fig. 1)

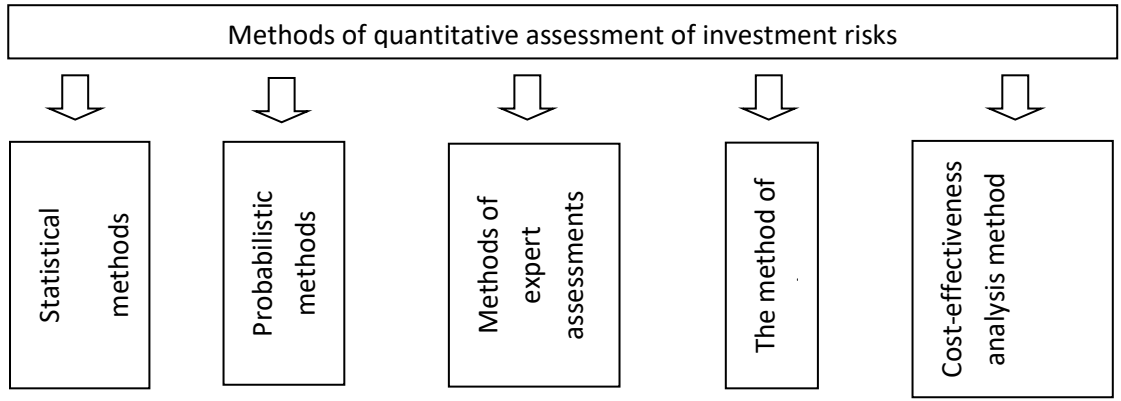

Fig.1 Methods of quantitative assessment of investment risks

The statistical method, based on the methods of mathematical statistics, has become widely popular for assessing the magnitude of investment risk. According to the statistical method, statistical data are analyzed. As a result of this analysis, the frequency of occurrence of certain events and the achievement of economic indicators are determined. Based on the values of economic indicators, their probabilistic forecast for the future is compiled. In the statistical approach, as a rule, for each of the possible options, the probability of occurrence of an event is determined. This allows you to predict possible outcomes based on the appropriate probabilities. Possible investment risks can be assessed using the coefficients of standard deviation, variance, and coefficient of variation [16]. The results of the influence of adverse events can be obtained based on an assessment of the mathematical expectations of the studied values. In our opinion, the main advantage of statistical methods is their simplicity, they do not require special mathematical training from the researcher. According to [17], the disadvantage of statistical methods is the need to use a large sample of data for reliable and objective analysis.

Let's say X is a random variable included in some probability space. Then the variance of this random variable can be written as [18]:

$$D[X] = M[(X - M(X))^2] \qquad (1)$$

In the future (1) will be called the variance relative to the average value. The arithmetic mean of the terms of a numerical series will be further referred to as the mathematical expectation relative to the average value $- M(X)$.

In practice, along with the variance, the standard deviation of a random variable is used [18]

$$\sigma = \sqrt{D[X]} \qquad (2)$$

In the future (2) will be called the standard deviation relative to the average value.

Using the coefficient of variation of a random variable, it is possible to determine the magnitude of the relative spread of this random variable [18]

$$V = \frac{\sigma}{M(X)} \qquad (3)$$

In the following (3) we will call the coefficient of variation of a random variable relative to the average value.

According to the rules $3\sigma$ [18], with a very high probability, the distribution of the random variable X lies in the interval $(M(X) - 3\sigma ; M(X) + 3\sigma)$. If the random variable X has a normal distribution, then the probability of the random variable X falling into the above-mentioned interval is 0.9973. Let's look at this with a specific example, define the interval $(M(X) - 3\sigma; M(X) + 3\sigma)$ for the inflation index in the Czech Republic in 2011-2021. At the same time, we will assume that the numerical series of values of the inflation index in the Czech Republic in 2011-2021 has a normal distribution.

Table1. Inflation in the Czech Republic 2011 – 2021 [19]

| Year number | 1 | 2 | 3 | 4 | 5 | 6 |
|---|---|---|---|---|---|---|
| Year | 2011 | 2012 | 2013 | 2014 | 2015 | 2016 |
| Inflation, % $I$ | 2,2 | 3,5 | 1,4 | 0,4 | 0,3 | 0,6 |
| Year number | 7 | 8 | 9 | 10 | 11 | |
| Year | 2017 | 2018 | 2019 | 2020 | 2021 | |
| Inflation, % $I$ | 2,4 | 2 | 2,6 | 3,3 | 3,3 | |

After simple calculations, we get

**σ** = 1,1265 , $M(I) = 2\%$ , **3σ** = 1,1265*3=3.3795 , $V(I)$=0,56325

$-3\sigma + M(I) \leq I \leq M(I) + 3\sigma ;\quad -1,3795 \leq M(I) \leq 5,3795$

where,

**σ** - the standard deviation of the inflation index relative to the average value;
$M(I)$ – the mathematical expectation of the inflation index relative to the average value;
$V(I)$- coefficient of variation of the inflation index relative to the average value;

Considering that we have made the assumption that the random value of the inflation index I has a normal distribution. The forecast that with a probability of 99.73% inflation will take a value within (-1.3795; 5.3795) will not be of practical use. This is equivalent to the fact that in the television news, during the weather forecast, we would be informed that tomorrow the air temperature would be in the range of -5 to 15 degrees Celsius with a 99% probability. Which, of course, is a very accurate prediction, from a mathematical point of view, but it is useless for practical use. Based on this forecast, we will not be able to choose clothes for our trip to work.

Thus, there is a need to find the variability of a random variable, which would be more suitable for practice in the case when the trend of a random variable differs from the linear form

## Methods

The Fermat-Torricelli point is a point on the plane from which the sum of the distances to the vertices of the triangle reaches a minimum value [20].

$$|AF| + |BF| + |CF| \to min \tag{4}$$

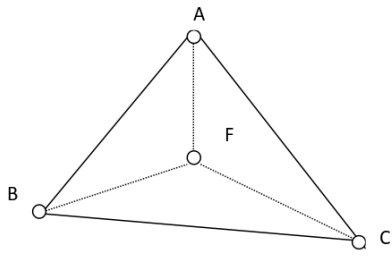

Fig 2. The Fermat-Torricelli point

Theorem [21] Suppose a triangle has all angless less then $\frac{2\pi}{3}$ the Fermat-Torricelli point has coordinates:

$$x = \frac{X}{2d\sqrt{3}} \;,\; y = \frac{Y}{2d\sqrt{3}} \;,\; \text{where } d = \frac{r_{12}^2 + r_{13}^2 + r_{23}^2}{2} + |S| \cdot \sqrt{3} \tag{5}$$

$$X = \sqrt{3}(x_1 r_{23}^2 + x_2 r_{13}^2 + x_3 r_{12}^2) + (x_1 + x_2 + x_3)|S|$$
$$+ sign(S)[(y_2 - y_1)(x_1 x_2 + y_1 y_2) +$$
$$+ (y_1 - y_3)(x_1 x_3 + y_1 y_3) + (y_1 - y_3)(x_1 x_3 + y_1) + (y_3 - y_2)(x_2 x_3 + y_2 y_3)$$

$$Y = \sqrt{3}(y_1 r_{23}^2 + y_2 r_{13}^2 + y_3 r_{12}^2) + (y_1 + y_2 + y_3)|S|$$
$$+ sign(S)[(x_2 - x_1)(x_1 x_2 + y_1 y_2) +$$
$$+ (x_1 - x_3)(x_1 x_3 + y_1 y_3) + (x_1 - x_3)(x_1 x_3 + y_1) + (y_3 - y_2)(x_2 x_3 + y_2 y_3)$$

$$S = x_1 y_2 + x_3 y_1 + x_2 y_3 - x_1 y_3 - x_2 y_1 - x_3 y_2$$

$r_1, r_2, r_3$ - the length of the side of the triangle , $x_1, x_2, x_3$ , $y_1, y_2, y_3$ – coordinates of the vertices of the triangel.

According to [22], a function can be called a certain relation between two sets X and Y if a single element from the set Y is assigned to each element of the set X.

Let's introduce the concept of the Fermat-Torricelli function.

To each point $\{x_2, x_3, \ldots, x_{m-1}\}$ let's match the corresponding Fermat-Torricelli point $\{\varphi(x_2), \varphi(x_3), \ldots, \varphi(x_{m-1})\}$. Obviously, for the extreme points $x_1$ and $x_m$ Fermat-Torricelli points do not exist.

Let's introduce the concept $F(X)$ variances with respect to Fermat-Torricelli points

$$F(X) = M(X - \varphi(X))^2 \tag{6}$$

Let's introduce the concept $S(X)$ standard deviation relative to Fermat-Torricelli points

$$S(X) = \sqrt{F(X)} = \sqrt{M(X - \varphi(X))^2} \tag{7}$$

Let's introduce the concept $M(X)$ the mathematical expectation of Fermat-Torricelli points

$$M(X) = \sum_{i=1}^{n} \frac{X_i - \varphi(X_i)}{n} \tag{8}$$

Let's introduce the concept $W(X)$ coefficient of variation of a random variable relative to Fermat-Torricelli points

$$W(X) = \frac{S(X)}{M(X - \varphi(X))} \tag{9}$$

The theorem.

**If M(X) – mathematical expectation with respect to Fermat-Torricelli points, S(X) – the standard deviation relative to the Fermat-Torricelli points, then $P(|X - M(X) \geq 4S(X)|) \leq 0,0625$**

For $a > 0$ inequality $P(|X - M(X) \geq a|)$
it is equivalent to inequality

$$P(|X - M(X)| \geq a) = P((X - M(X)^2 \geq a^2) \leq \frac{M(X - \varphi(X))^2}{a^2} = \frac{F(X)}{a^2}$$

$a = 4 S(X)$ ;

$$P(|X - M(X) > 4 \cdot S(X)|) \leq \frac{F(X)}{4^2 \cdot (S(X))^2} = \frac{1}{16} = 0,0625$$

Let's formulate a rule 4 S.

For any random variable X, the probability that its value will deviate from the mathematical expectation with respect to Fermat-Torricelli points M(X) by at least four standard deviations with respect to Fermat-Torricelli points, but not more than 0.0625.

*That is, regardless of whether the random variable has a normal distribution or not. The probability of finding this random variable within*
$$-4S(X) + M(X) \leq M(X) \leq M(X) + 4S(X)$$
*equal to 93,75%.*

According to the author, this is a fairly acceptable approximation of a random variable from a practical point of view. As an example, let's do calculations for the data Table 1.

$$|I - \varphi(I)| = \sqrt{(\text{Year number} - \varphi(\text{Year number}))^2 + (I - \varphi(I))^2}$$

2: $|I - \varphi(I)| = \sqrt{(2 - 1{,}7912)^2 + (3{,}5 - 2{,}4661)^2} = \sqrt{0.04359744 + 1.06894921} = 1{,}0548$

3: $|I - \varphi(I)| = \sqrt{(3 - 3)^2 + (1{,}4 - 1{,}4)^2} = 0$

4: $|I - \varphi(I)| = \sqrt{(4 - 4)^2 + (0{,}4 - 0{,}4)^2} = 0$

5: $|I - \varphi(I)| = \sqrt{(5 - 5)^2 + (0{,}3 - 0{,}3)^2} = 0$

6: $|I - \varphi(I)| = \sqrt{(6 - 6)^2 + (0{,}6 - 0{,}6)^2} = 0$

7: $|I - \varphi(I)| = \sqrt{(7 - 7{,}1264)^2 + (2{,}4 - 2{,}1045)^2} = \sqrt{0{,}01597696 + 0{,}08732025} = 0{,}3214$

8: $|I - \varphi(I)| = \sqrt{(8 - 8)^2 + (2 - 2)^2} = 0$

9: $|I - \varphi(I)| = \sqrt{(9 - 9)^2 + (2{,}6 - 2{,}6)^2} = 0$

10: $|I - \varphi(I)| = \sqrt{(10 - 10)^2 + (3{,}3 - 3{,}3)^2} = 0$

11: $|I - \varphi(I)| = \sqrt{(11 - 10{,}6380)^2 + (3{,}3 - 3{,}5719)^2} = \sqrt{0{,}131044 + 0{,}07392961} = 0{,}4527$

Table 2. Coordinates of Fermat-Torricelli points for the inflation chart in the Czech Republic 2011-2021 (calculated by the author himself)

| Year number | 1,7912 | 3 | 4 | 5 | 6 |
|---|---|---|---|---|---|
| Year | 2011,7912 | 2013 | 2014 | 2015 | 2016 |
| Inflation, % $I$ | 3,5 | 1,4 | 0,4 | 0,3 | 0,6 |
| Inflation, % $\varphi(I)$ | 2,4661 | 1,4 | 0,4 | 0,3 | 0,6 |
| $|I - \varphi(I)|$, % | 1,3795 | 0 | 0 | 0 | 0 |

| Year number | 7,1264 | 8 | 9 | 10 | 10,6380 |
|---|---|---|---|---|---|
| Year | 2017,1264 | 2018 | 2019 | 2020 | 2020,6380 |
| Inflation, % $I$ | 2,4 | 2 | 2,6 | 3,3 | 3,3 |
| Inflation, % $\varphi(I)$ | 2,1045 | 2 | 2,6 | 3,3 | 3,5719 |
| $|I - \varphi(I)|$, % | 0,3045 | 0 | 0 | 0 | 0,2719 |

$$M(\varphi(I)) = \frac{2,4661 + 1,4 + 0,4 + 0,3 + 0,6 + 2,1045 + 2 + 2,6 + 3,3 + 3,5719}{10} = 1,87425$$

After simple calculations, we get the final result:

$S(I) = 0.35, M(\varphi(I)) = 1,87\%$ ; $4\,S(I) = 1,4$ ;

$-4S(I) + \varphi(I) \leq I \leq \varphi(I) + 4S(I)$ ; $-1,4 + 1,87 \leq I \leq 1,87 + 1,4$ ;

$0,47 \leq \varphi(I) \leq 3,27; W(I) = \frac{0,35}{1,87} = 0,187$

$S(I)$ - the standard deviation of the inflation index relative to the Fermat-Torricelli points;

$M(I)$ – mathematical expectation of the inflation index relative to Fermat-Torricelli points;

$W(I)$- coefficient of variation of the inflation index relative to Fermat-Torricelli points;

$\varphi(I) -$ the inflation index according to a curve that interpolates Fermat-Torricelli points.

The Fermat-Torricelli point interpolation function I (10) for the inflation index in the Czech Republic in the years (2011-2021) was constructed by the author in the work [23]

I =(0.264901377876643)*exp((0.249672956416996)*t)+(-0.007782663831297 +     (10)
0.015129431149835i)*exp((0.076090999247734 + 2.511250329378980i)*t)+(-0.007782663831297 -
0.015129431149835i)*exp((0.076090999247734 - 2.511250329378980i)*t)+(-1.671150941557596 -
0.869131660330525i)*exp((-0.303576461438207 + 1.138618581934044i)*t)+(-1.671150941557596 +
0.869131660330525i)*exp((-0.303576461438207 - 1.138618581934044i)*t)+(-
0.014659833689592)*exp((-0.249672956416996)*t)+(0.138184785734736 -
0.180858361988375i)*exp((-0.076090999247734 - 2.511250329378980i)*t)+(0.138184785734736 +
0.180858361988375i)*exp((-0.076090999247734 + 2.511250329378980i)*t)+(0.003015325281862 -
0.001937822578385i)*exp((0.303576461438207 - 1.138618581934044i)*t)+(0.003015325281862 +
0.001937822578385i)*exp((0.303576461438207 + 1.138618581934044i)*t)

Function graph (10) It was also built by the author in the work [23] (Fig.3).

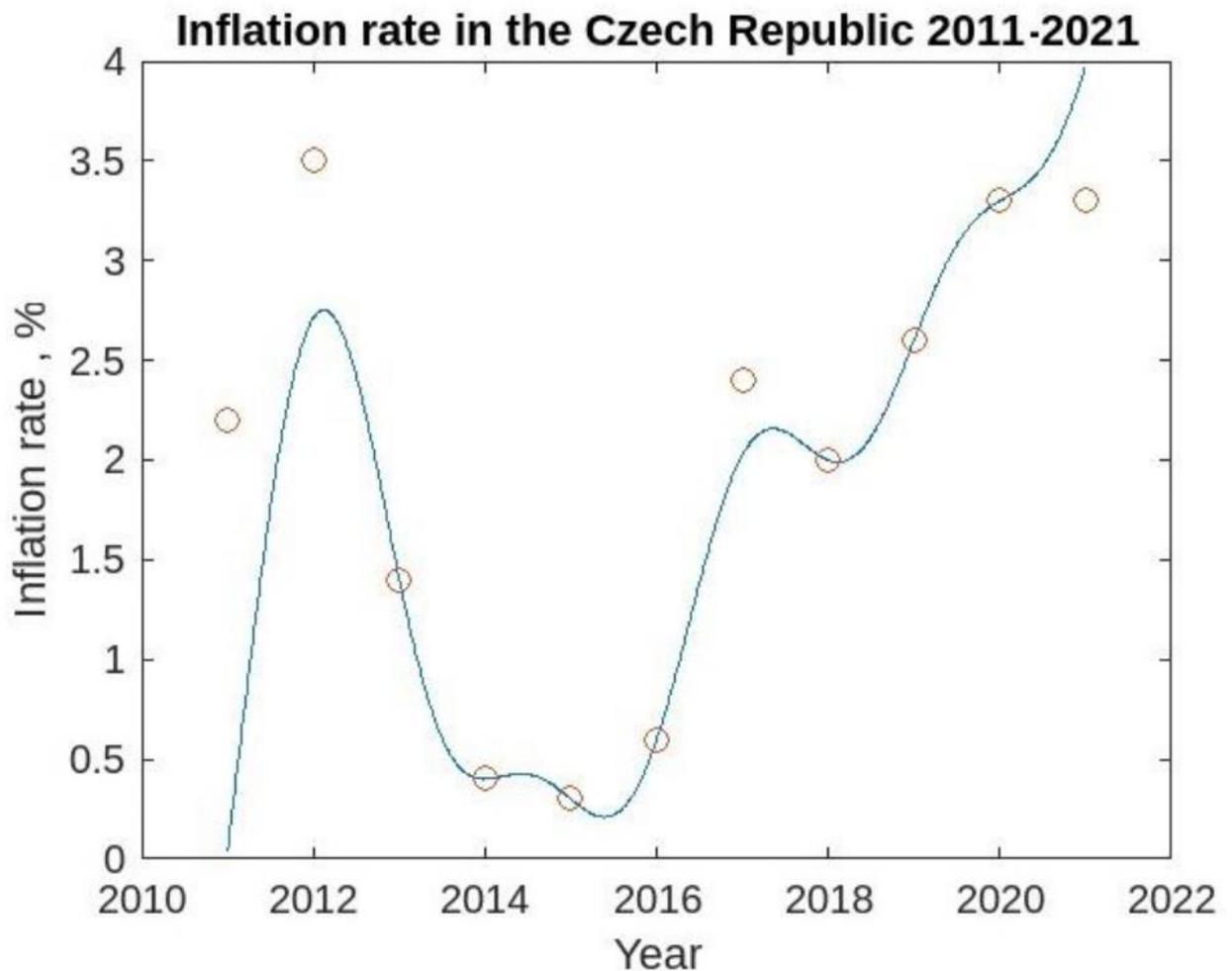

Fig.3. The inflation index of the Czech Republic in 2010-2021 was built by the author [23].

where   o - Statistical data on inflation in the Czech Republic for 2011-2021 [19]

▬▬▬ -interpolation of Fermat-Torricelli points using the function .

## Discussion

Let's write down the results of the projected value of the inflation index in the Czech Republic based on statistical data [19] 2011-2021 (Table 3)

Tabl.3 Comparison of the forecast value of the inflation index in the Czech Republic in 2011-2021 calculated by the traditional method and the method that uses Fermat-Torricelli points (calculated by the author independently)

| The traditional method | | | $-1,3795 \leq M(I) \leq 5,3795$ |
|---|---|---|---|
| σ = 1,1265 | $M(I) = 2\%$ | 3σ = 3,3795 | $V(I)=0,56325$ |
| Fermat-Torricelli point method | | | $0,47 \leq \varphi(I) \leq 3.27$ |
| S = 0,35 | $M(\varphi(I)) = 1,87\%$ | 4S = 1,4 | $W(I) = 0,187$ |

The traditional method of estimating the forecast value of the inflation index in the Czech Republic in 2011-2021 , which uses σ - the standard deviation relative to the average value for the inflation index (1) , $M(I)$ – the mathematical expectation of the inflation index relative to the average value (2) , $V(I)$ – coefficient of variation relative to the average value (3) (assuming that $I_k$ $k = 1, ...,10$ has a normal distribution) then, with probability 99,73% inflation will take a value within $(-1,3795; 5,3795)$ %.

The method of estimating the forecast value of the inflation index in the Czech Republic in 2011-2021 , which uses, $S(I)$ standard deviation relative to Fermat-Torricelli points (7) , $M(I)$ mathematical expectation with respect to Fermat-Torricelli points mathematical expectation with respect to Fermat-Torricelli points (8) , $W(I)$ coefficient of variation of a random variable relative to Fermat-Torricelli points (9) (assuming that the numerical series $I_k$ $k = 1, ...,10$ there is no normal distribution), then the estimation method based on the use of Fermat-Torricelli points predicts that with probability 93,75% inflation will take a value within (0,47; 3,27) %.

Although mathematically the forecast of the inflation index is likely 99,73% , that inflation will take on a value within $(-1,3795; 5,3795)$ % more accurate than a forecast with probability 93,75% inflation will take a value within (0,47; 3,27) %.
However, for practical use, the second forecast will be better than the first one for the following reasons:

1) The value of the inflation index near the values of -1.3% and 5.3% is very unlikely in reality. In this regard, it is inappropriate to take this possibility into account in practical calculations, according to the author.
2) The forecast of the inflation index based on the use of Fermat-Torricelli points in the range of (0.47; 3.27)%, according to the author, is closer to the real situation. It gives a narrower range of possible values of the inflation index.
3) In any case, statistical forecasts cannot give a forecast close to 100%, since statistical forecasts, as a rule, do not take into account the possibility of very unlikely factors, but having a great influence on a random variable (natural disasters, wars, pandemics, etc.)

## Conclusions

Linear dependencies are very rare in nature. The use of Fermat-Torricelli points can be an effective mathematical tool for analyzing numerical series that have a large variance, a pronounced nonlinear trend, or do not have a normal distribution of a random variable.

Smoothing numerical series by constructing Fermat-Torricelli points reduces the influence of the random component on the final result.

The presence of a normal distribution of a random variable for numerical series that relate to long time intervals is an exception to the rule rather than an axiom. The external environment (international economic relations, scientific and technological progress, political events) is constantly changing, which in turn, in general, does not give grounds to assert that under these conditions a random variable satisfies the requirements of the Gauss-Markov theorem.